\documentclass[usenatbib]{mnras}

\usepackage{graphicx}
\title[Inner heliosphere solar wind categorisation]{Diagnosing solar wind origins using in-situ measurements in the inner heliosphere}
\author[D. Stansby et al.]{
D. Stansby,$^{1}$\thanks{E-mail: \href{david.stansby14@imperial.ac.uk}{david.stansby14@imperial.ac.uk}}
T. S. Horbury,$^{1}$
and L. Matteini$^{2}$
\\
$^{1}$Department of Physics, Imperial College London, London, SW7 2AZ, UK\\
$^{2}$LESIA, Observatoire de Paris, Universit\'e PSL, CNRS, Sorbonne Universit\'e, \\~~Univ. Paris Diderot, Sorbonne Paris Cit\'e, 5 place Jules Janssen, 92195 Meudon, France}
\pubyear{2018}
\begin{document}
\maketitle

% Start of abstract
\begin{abstract}
Robustly identifying the solar sources of individual packets of solar wind measured in interplanetary space remains an open problem. We set out to see if this problem is easier to tackle using solar wind measurements closer to the Sun than 1 AU, where the mixing and dynamical interaction of different solar wind streams is reduced. Using measurements from the Helios mission, we examined how the proton core temperature anisotropy and cross helicity varied with distance. At 0.3 AU there are two clearly separated anisotropic and isotropic populations of solar wind, that are not distinguishable at 1 AU. The anisotropic population is always Alfv\'enic and spans a wide range of speeds. In contrast the isotropic population has slow speeds, and contains a mix of Alfv\'enic wind with constant mass fluxes, and non-Alfv\'enic wind with large and highly varying mass fluxes. We split the in-situ measurements into three categories according these observations, and suggest that these categories correspond to wind that originated in the core of coronal holes, in or near active regions or the edges of coronal holes, and as small transients form streamers or pseudostreamers. Although our method by itself is simplistic, it provides a new tool that can be used in combination with other methods for identifying the sources of solar wind measured by Parker Solar Probe and Solar Orbiter.
\end{abstract}

\begin{keywords}
Sun: heliosphere -- solar wind.
\end{keywords}

\section{Introduction}
The solar wind is a continuous flow of plasma, travelling away from the Sun to fill interplanetary space. Although well studied by both in-situ and remote sensing instruments, robustly determining the solar origin of all the solar wind measured in-situ by spacecraft is still an open problem. Excluding large ejecta, the solar wind has traditionally been classified according to the average speed of protons, which make up $\sim$95--99\% of the wind by ion number density. It is well known that the solar source of fast solar wind ($v >\sim$ 500 km/s) is open field lines rooted in coronal holes on the surface of the sun \citep{Krieger1973, Sheeley1976, Cranmer2009}, and that it is comprised of a slowly varying bulk speed superposed with shorter timescale Alfv\'enic velocity fluctuations above this background level \citep{Belcher1971a, Thieme1989, Matteini2015}. The Alfv\'enic fluctuations are believed to be generated close to the surface of the Sun where the wind is sub-Alfv\'enic, and then propagate outwards into the heliosphere on long-lived open magnetic field lines \citep{Belcher1971a, Cranmer2005}.  As well as the global dynamics, the local kinetic properties of fast solar wind are also well known: the protons in fast solar wind have large temperature anisotropies and a low plasma beta in the inner heliosphere \citep{Marsch1982b, Marsch2004, Matteini2007}. In addition, alpha particles, which constitute 1\% -- 5\% of the solar wind by ion number density, exhibit large magnetic field aligned drifts relative to the protons in the fast solar wind \citep{Marsch1982c, Steinberg1996}.

In contrast to the fast wind, the plasma properties of the slow solar wind ($v <\sim$ 500 km/s) are much more variable \citep{Schwenn2006}, and although it must have a number of different solar sources \citep{Abbo2016}, it is not clear how these sources contribute to the different parts of the slow solar wind measured in-situ. In general there are two possible generation mechanisms for the slow wind: it can flow continuously on magnetic field lines that maintain a connection from the base of the corona to the heliosphere, in a similar manner to the fast solar wind \citep{Wang1990, Cranmer2007a, Wang2010}, or can be released transiently from closed magnetic field lines undergoing interchange reconnection with the open magnetic field lines that connect to the heliosphere \citep{SheeleyJr.1997, Einaudi2001, Rouillard2010, Higginson2017a}.

During initial analysis of Helios data, \cite{Marsch1981} discovered a portion of slow solar wind measured at 0.3 AU that, apart from its speed, had the same properties of fast solar wind: large proton-alpha drift speeds, large proton core temperature anisotropies, and highly Alfv\'enic wave activity. In addition, \cite{Roberts1987a} described an 80 day interval in the Helios data where the purest Alfv\'enic fluctuations were in slow solar wind. The strong Alfv\'enic wave activity during these periods implies that the wind was released on open magnetic field lines, allowing the Alfv\'en waves to freely propagate outwards from the corona to the point of measurement. More recently the presence of an ``Alfv\'enic slow wind'' has been studied statistically using data taken at 1 AU, and independent of solar activity the slow solar wind is comprised of both non-Alfv\'enic and Alfv\'enic components \citep{DAmicis2015a}. These results hint that some slow solar wind has exactly the same source (and therefore properties) as fast solar wind, but just happens to be released at a slower speed.

As well as protons and alphas, much less abundant heavy ions are measured in the solar wind, which can be used as a more direct proxy for solar source than the proton speed. As the solar wind travels away from the sun it effectively becomes collision-less within a few solar radii \citep{Hundhausen1968}. This means that ions are no longer able to gain or lose electrons through electron-ion collisions, and the fraction of different charge states becomes frozen in. Ion charge state ratios therefore act as a tracer of the plasma properties at the freezing in point. The most commonly used ratios are O$^{7+}$/O$^{6+}$ and C$^{6+}$/C$^{4+}$, which are positively correlated with the electron temperature at the freezing in point \citep{Hundhausen1968, Bochsler2007, Landi2012}. Low charge state ratios are present in wind that originates in coronal holes, which have relatively low electron temperatures, and high charge state ratios are present in streamer belt plasma that has relatively high electron temperatures. This information can therefore be used to distinguish between coronal hole and non-coronal hole solar wind \citep{Geiss1995, Zhao2009}. As an example, the Alfv\'enic slow wind identified by \cite{DAmicis2015a} had similar low charge state ratios to the fast solar wind, indicating it had similar solar origins \citep{DAmicis2016}, further reinforcing the need to go beyond classifying solar wind based solely on the average proton speed.

Although the Helios mission was equipped with an instrument for measuring heavy ions in the solar wind \citep{Rosenbauer2018}, it is believed that the data from this instrument has been lost. However, the case studies of \cite{DAmicis2015a} and \cite{DAmicis2016} hint that when heavy ion measurements are not available the Alfv\'enicity of the solar wind fluctuations may be a more reliable proxy for solar wind origin than speed. The problem with using Alfv\'enicity as a categorisation variable is that the solar wind becomes systematically less Alfv\'enic with distance \citep{Roberts1987a, Bruno2007, Iovieno2016}, due to both small scale turbulent evolution and large scale velocity shears and interaction regions \citep{Bruno2006}. This means not all solar wind that started off Alfv\'enic near the sun is still Alfv\'enic when it is measured at 1~AU. In this paper we mitigate this problem by using the unique Helios data, with measurements of solar wind plasma from 0.3 AU to 1 AU, to link properties measured in-situ, that are only observable at distances $<$ 1 AU, to solar sources. The Helios data are described in section \ref{sec:data}, and statistical results presented in section \ref{sec:results}. Based on these observations we construct three different categories of solar wind, and in section \ref{sec:solar sources} we place possible solar sources in to each of these categories. In section \ref{sec:compare} we compare our categorisation scheme with other schemes, and then conclude and present a set of predictions for the upcoming Solar Orbiter and Parker Solar Probe missions in section \ref{sec:predictions}.

%%%%
\section{Data}
\label{sec:data}
The data used here were measured by the twin Helios spacecraft, which were operational during the late 1970s and early 1980s. Both spacecraft had an electrostatic analyser for measuring the ion distribution function at 40.5 second cadence \citep{Schwenn1975}, and two different fluxgate magnetometers for measuring the magnetic field \citep{Musmann1975, Scearce1975}. Here we use a re-analysis of the ion distribution functions which fitted a bi-Maxwellian to the proton core population present in the experimentally measured ion distribution functions. This dataset provides the proton core number density ($n_{p}$), velocity ($\mathbf{v}_{p}$), temperatures parallel ($T_{p \parallel}$) and perpendicular ($T_{p \perp}$) to the magnetic field, and corresponding magnetic field values ($\mathbf{B}$) at a maximum cadence of 40.5 seconds. For more details on the fitting procedure and access to the data see \cite{Stansby2018e}. Note that all the data presented in this article are parameters of the proton core population, and not numerical moments of the overall ion distribution. The total temperature was calculated as $T_{p}=\left (2T_{p\perp} + T_{p\parallel} \right ) / 3$, the temperature anisotropy as $T_{p\perp} / T_{p\parallel}$, the parallel plasma beta as $\beta_{p} = 2\mu_{0}n_{p} k_{B} T_{p\parallel} / \left | \mathbf{B} \right |^{2}$, and the Alfv\'en speed as $v_{A} = \left | \mathbf{B} \right | /\sqrt{n_{p} m_{p} \mu_{0}}$.

To avoid contamination of very large transients all of the intervals listed as coronal mass ejections by \cite{Liu2005a} were removed from the dataset before further analysis. The state of the Sun undergoes an 11 year solar cycle, oscillating between solar minimum and solar maximum. Because the highest quality Helios data were taken early in the mission, only data taken in the years 1974 -- 1978 inclusive (during the solar minimum between cycles 20 and 21) were used.

\subsection{Alfv\'enicity}
In order to classify the solar wind as Alfv\'enic or non-Alfv\'enic, the cross helicity was calculated in every 20 minute interval where at least 10 velocity and magnetic field data points were available. The cross helicity is defined as
\begin{equation}
	\sigma_{c} = 2 \frac{\left \langle \mathbf{v} \cdot \mathbf{b} \right \rangle}{\left \langle \left | \mathbf{v} \right |^{2} + \left | \mathbf{b} \right |^{2} \right \rangle}
\end{equation}
where $\mathbf{v} = \mathbf{v}_{p} - \mathbf{v}_{p0}$ are the proton velocity fluctuations in the wave frame, $\mathbf{v}_{p0}$ is the local Alfv\'en wave phase velocity, $\mathbf{b} = v_{A} \left ( \mathbf{B} / \left | \mathbf{B} \right | \right )$ is the magnetic field in velocity units, and $\left \langle \right \rangle$ indicates a time average over all points in a 20 minute interval \citep{Bruno2013a}. $\mathbf{v}_{p0}$ was calculated using the method given by \cite{Sonnerup1987}, which finds the local de-Hoffman Teller frame of reference in which $\left \langle \left | \mathbf{v} \times \mathbf{b} \right |^{2} \right \rangle$ is minimised; by construction, this is the value of $\mathbf{v}_{p0}$ for which the absolute value of $\sigma_{c}$ is maximised. Although a plasma with Alfv\'en waves propagating in opposite directions can have low values of $\sigma_{c}$, in this paper ``Alfv\'enic" is specifically reserved to denote a plasma where Alfv\'en waves propagate predominantly in only one direction.

The magnitude of $\sigma_{c}$ indicates whether the fluctuations in the plasma are predominantly uni-directional Alfv\'en waves ($\left | \sigma_{c} \right | \approx 1$) or not ($\left | \sigma_{c} \right | < 1$). For Alfv\'enic periods, the sign of $\sigma_{c}$ determines the direction of wave propagation with respect to the local magnetic field. Because Alfv\'en waves in the solar wind almost always travel away from the Sun \citep{Gosling2009a}, the sign of $\sigma_{c}$ is a reliable proxy for the magnetic polarity of the solar wind. 

\subsection{Entropy}
Heavy ion charge state data measured at 1 AU is commonly used to diagnose the solar origin of solar wind. Unfortunately there is no heavy ion data available from the Helios mission, so instead proton specific entropy was used as a proxy. The specific entropy argument is easily calculated from the proton distribution parameters and given by
\begin{equation}
	S_{p} = T_{p} n_{p}^{\alpha - 1}
	\label{eq:entropy}
\end{equation}
where $\alpha$ is the polytropic index of the fluid. Here $\alpha$ was taken to be 1.5, the value used by \cite{Pagel2004} and \cite{Stakhiv2016} who studied correlations between entropy and composition, and whose results are used in section \ref{sec:composition} to make an indirect link between proton temperature anisotropy and heavy ion charge states using $S_{p}$ as an intermediate variable.

%%%%
\section{Results}
\label{sec:results}
%%%
\subsection{Statistics}
\label{sec:stats}

\begin{figure}
	\includegraphics[width=\columnwidth]{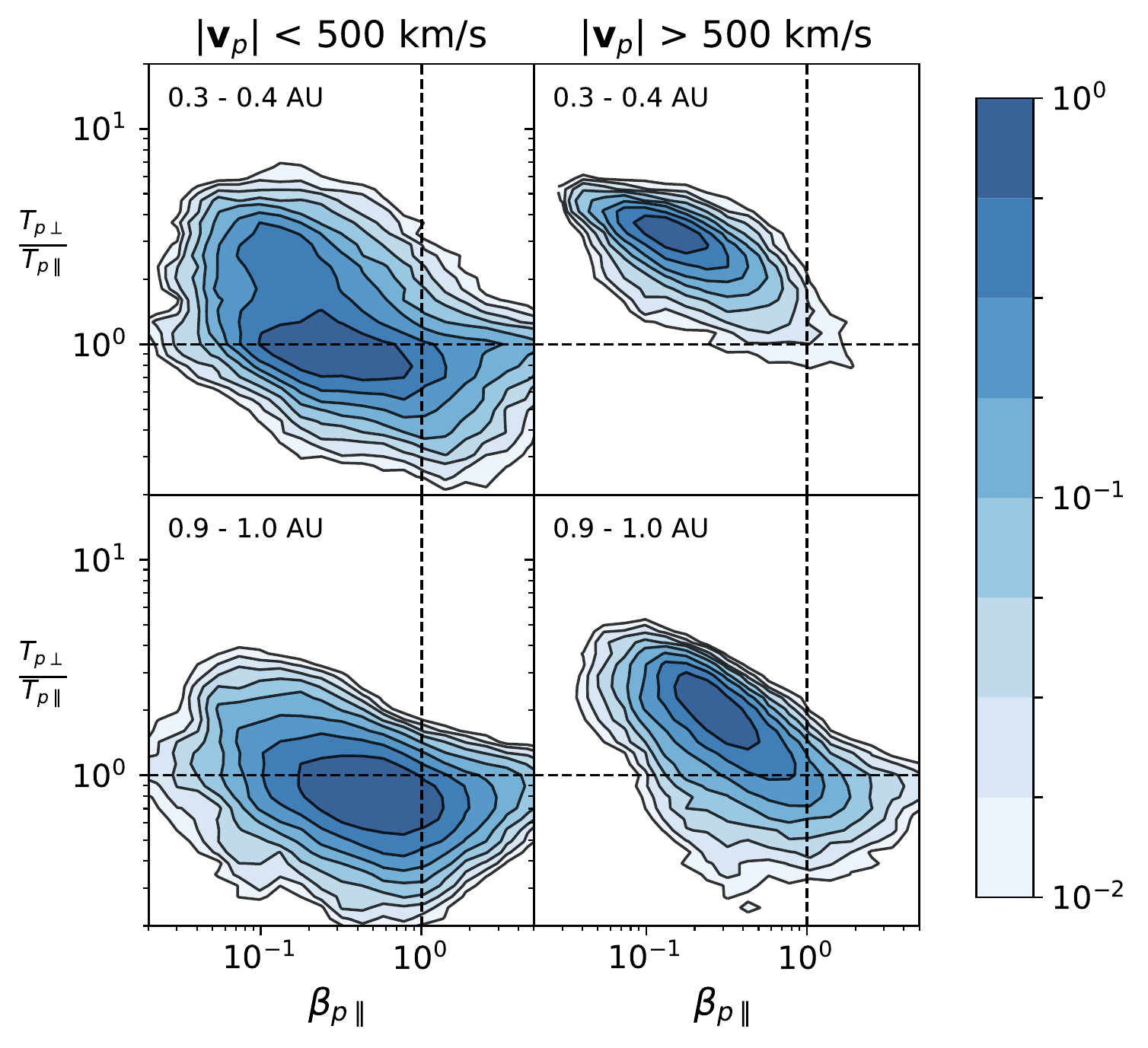}
	\caption{Distribution of data in the $\beta_{p\parallel}$ - $T_{p\perp} / T_{p\parallel}$ plane. Contours are interpolated from 2D histogram counts with 40 bins logarithmically spaced along each axis and normalised to the maximum bin count. Left panels shows slow solar wind and right panels show fast solar wind. Top panels show the distribution at 0.3 -- 0.4 AU and bottom panels show 0.9 -- 1.0 AU. }
	\label{fig:ani beta}
\end{figure}

Figure \ref{fig:ani beta} shows the distribution of the solar wind in the $\beta_{p\parallel}-T_{p\perp} / T_{p\parallel}$ plane measured by Helios at heliocentric distances of 0.3 -- 0.4 AU and 0.9 -- 1.0 AU, and split into slow and fast wind using a simple cut in speed. The distribution at 0.3 AU has previously been presented for an individual high speed stream by \cite{Matteini2007}. The top right panel of figure \ref{fig:ani beta} shows that the $\sim$100-hour continuous high speed stream in \cite{Matteini2007} is representative of all 542 hours of fast solar wind measured by Helios at 0.3 AU during solar minimum. The distribution at 1 AU is also well known, and the Helios data measured in the years 1974 -- 1978 (bottom two panels of figure \ref{fig:ani beta}) agrees well with data from the WIND spacecraft measured from 1995 to 2012 \citep[e.g.][]{Maruca2011}. As the wind propagates outwards the protons become more isotropic and the plasma beta increases, primarily due to adiabatic evolution \citep{Chew1956, Matteini2011}, although by 1 AU the fast wind protons have not yet reached the equilibrium state of $T_{p\perp} / T_{p\parallel} = 1$. In contrast, at 1 AU the slow solar wind is distributed around $T_{p\perp} / T_{p\parallel} = 1$, where it is thought to be maintained during transit by a combination of collisions and kinetic instabilities that are active when $\beta_{p \parallel} \geq 1$ \citep[e.g.][]{Kasper2002, Hellinger2006a, Bale2009, Yoon2016e}.

\begin{figure}
	\includegraphics[width=\columnwidth]{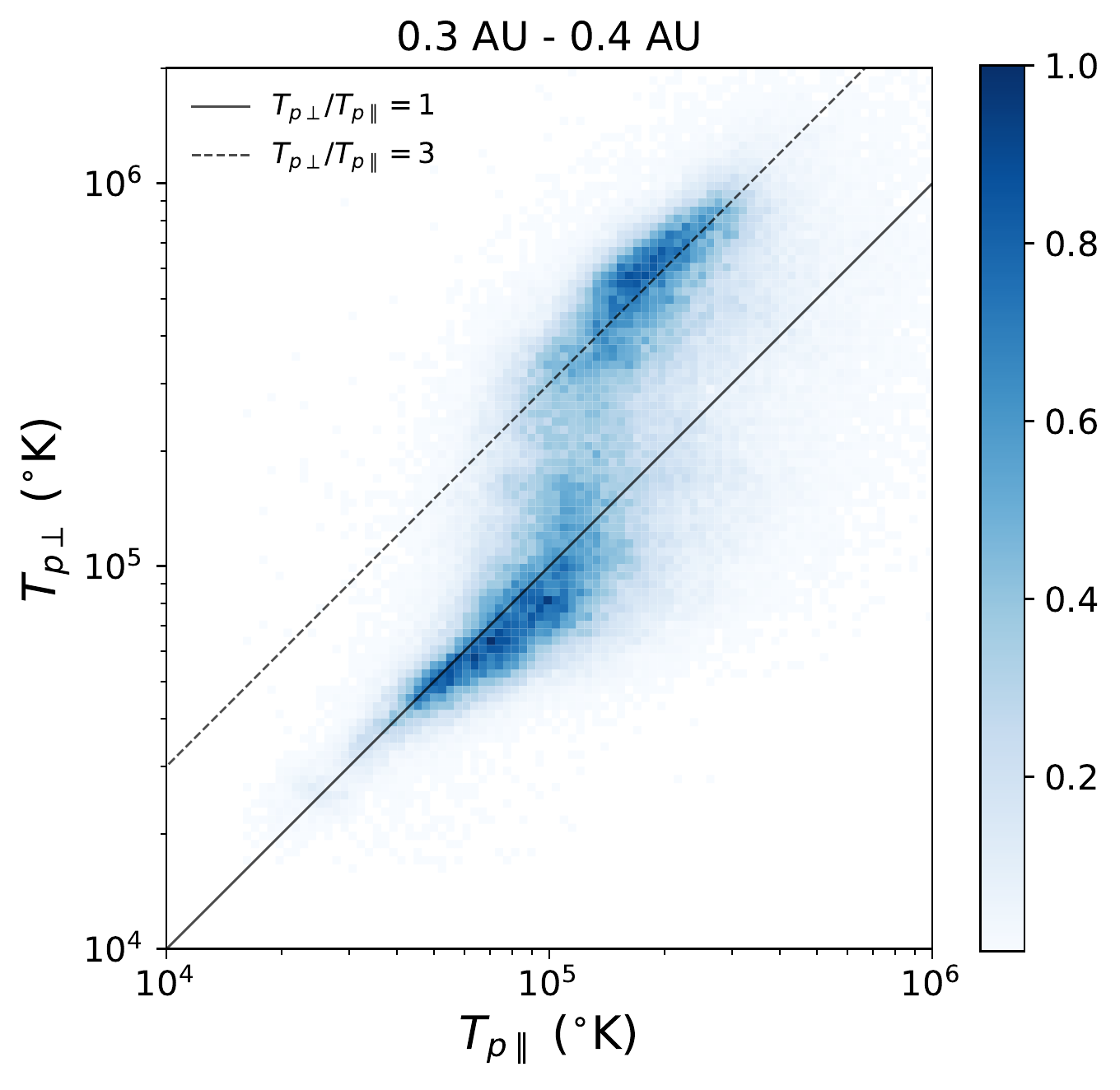}
	\caption{Joint probability distribution of parallel and perpendicular proton temperatures at 0.3 -- 0.4 AU. Histogram values are bin counts normalised to the maximum bin value. In this parameter space lines of constant temperature anisotropy are diagonal with a gradient of 1, with two examples shown for reference.}
	\label{fig:temp hists}
\end{figure}

At 0.3 AU the majority of the slow solar wind is also spread around $T_{p\perp} / T_{p\parallel} = 1$, but there is a significant fraction with $T_{p\perp} / T_{p\parallel} > 2$ and $\beta_{p} \ll 1$ (figure \ref{fig:ani beta} top left). This is the same region in the parameter space that fast solar wind occupies at 0.3 AU, which implies that there is a portion of slow solar wind that has the same kinetic properties as fast solar wind.  Instead of partitioning the data by speed, figure \ref{fig:temp hists} shows the joint distribution of $T_{p\perp}$ and $T_{p\parallel}$ for all measurements between 0.3 AU and 0.4 AU. In this parameter space two populations are clearly distinct: one centred around $T_{p\perp} / T_{p\parallel} = 1$ and one centred around $T_{p\perp} / T_{p\parallel} = 3$. All fast solar wind occupies the anisotropic population, but the slow wind is split between the two populations (as shown in figure \ref{fig:ani beta}).

\begin{figure}
	\centering
	\includegraphics[width=\columnwidth]{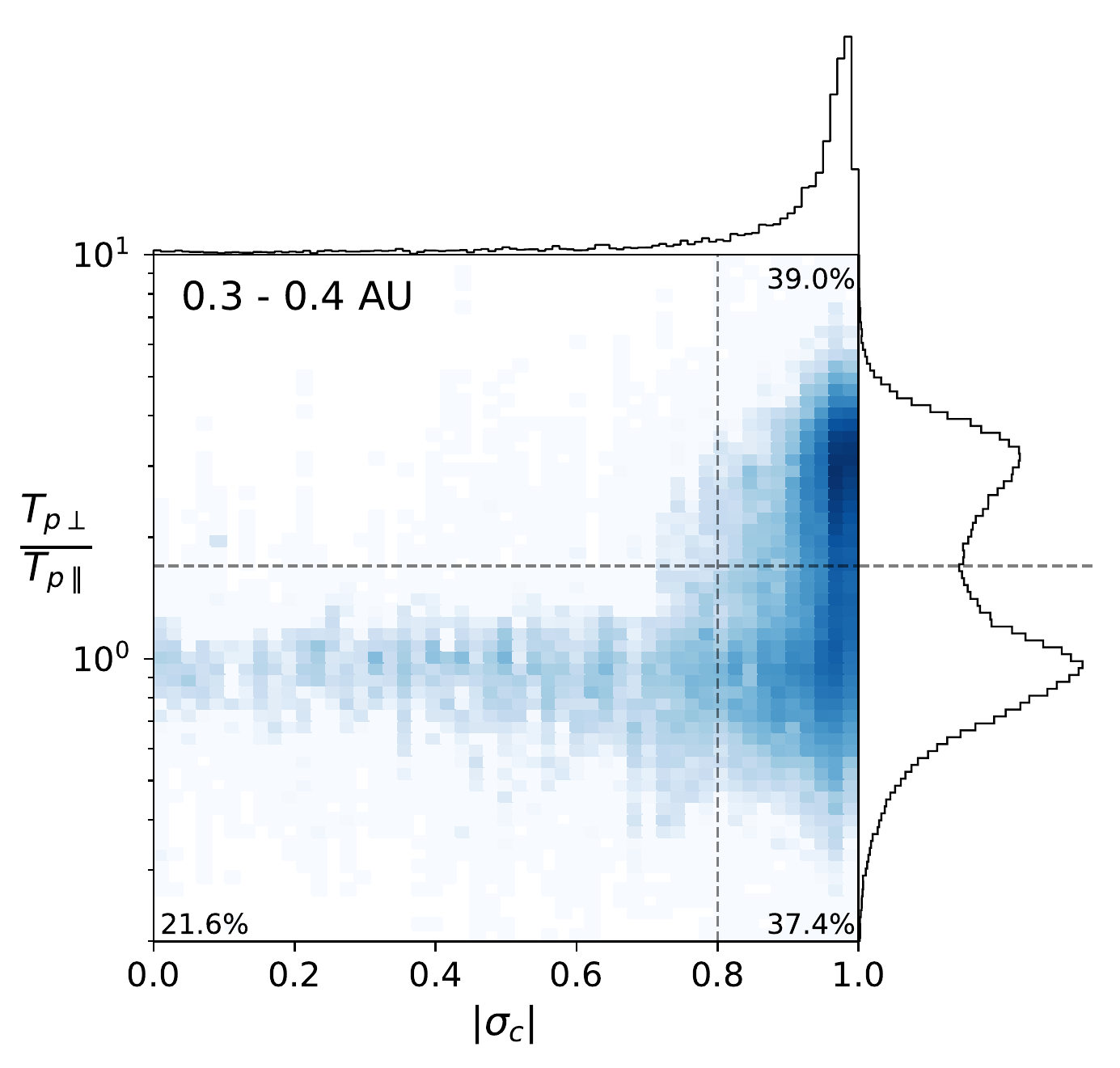}
	\includegraphics[width=\columnwidth]{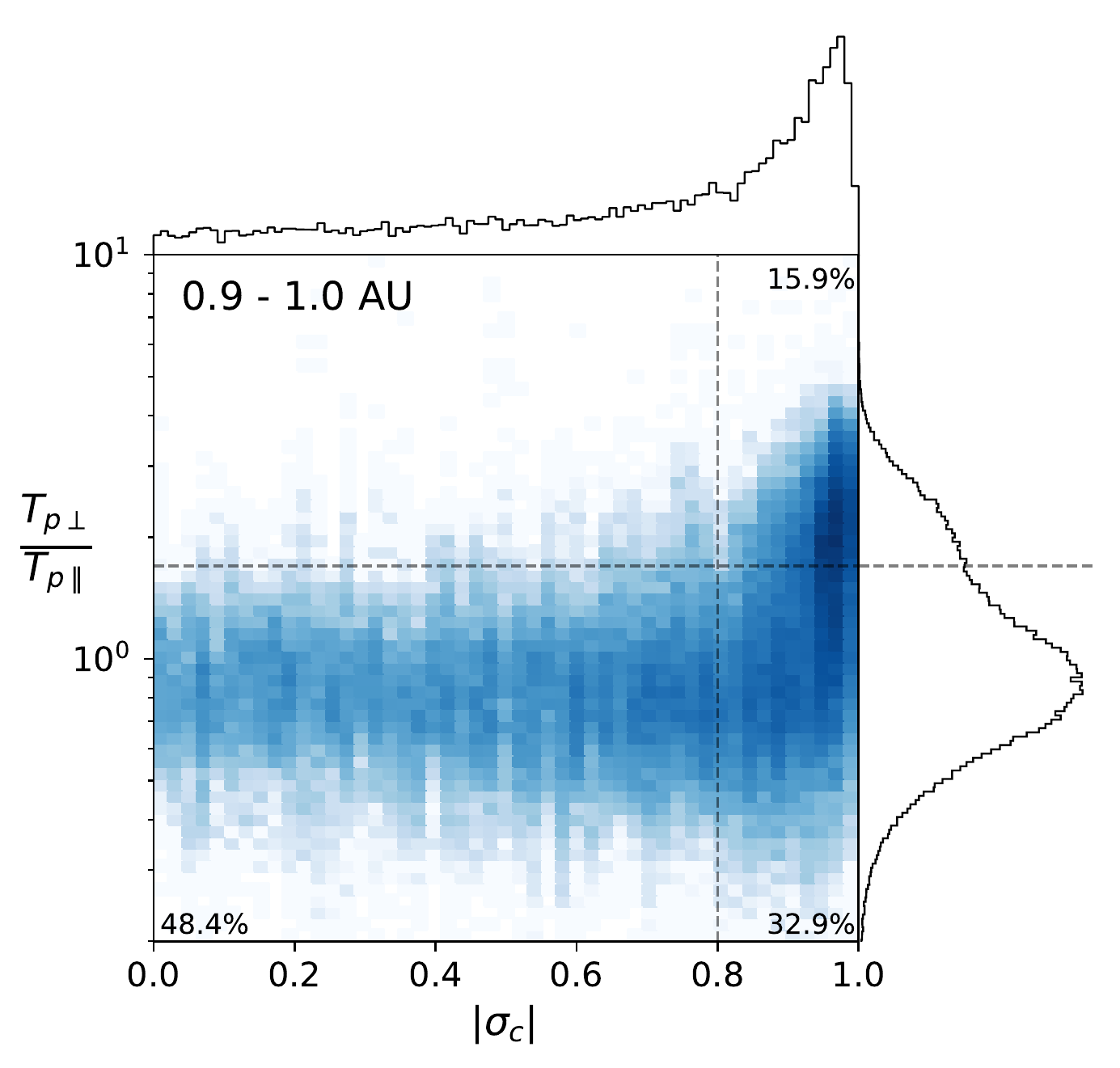}
	\includegraphics[width=0.7\columnwidth]{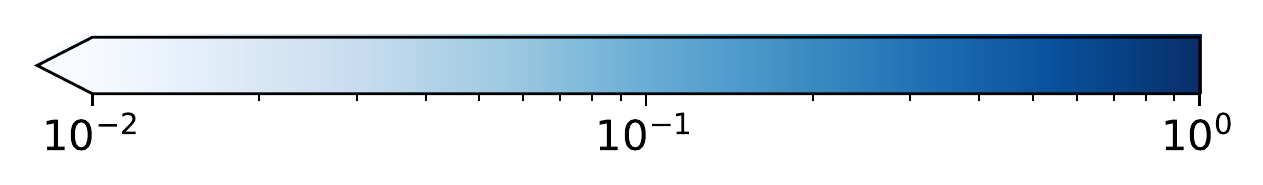}
	\caption{2D histograms of temperature anisotropy against absolute cross helicity (main panels) with adjoining 1D histograms of temperature anisotropy (right panels) and cross helicity (top panels). 2D bin counts are normalised to the maximum bin count. 1D histograms are linearly scaled. The horizontal and vertical dashed lines show the partitioning of the parameter space into three distinct regions; see text in section \ref{sec:stats} for more details. Percentages indicate the fraction of data points in each of the three regions.}
	\label{fig:alfven hists}
\end{figure}

As well as being anisotropic, the fast solar wind is filled by anti-sunward propagating Alfv\'en waves. This is a statement about the global dynamics of the plasma, in contrast to the local kinetic properties given by the parallel and perpendicular temperatures, and is a consequence of the wind being released on long-lived open field lines. To investigate whether the anisotropic wind as a whole has the same Alfv\'enic property as the fast solar wind, figure \ref{fig:alfven hists} shows the joint probability distribution of temperature anisotropy and cross helicity at both 0.3 AU and 1 AU. The distribution of temperature anisotropy is clearly bi-modal at 0.3 AU with a minimum at $T_{p\perp} / T_{p\parallel} = 1.7$, a feature that is not observable at 1 AU due to the 0.3 AU anisotropic component becoming more isotropic with radial distance (as a result of adiabatic evolution). The distribution stops being clearly bi-modal at radial distances greater than around 0.8 AU (not shown). In the rest of this paper data taken at 0.3 AU to 0.4 AU are presented, but the qualitative properties discussed are present at all radial distances from 0.3 AU to 0.8 AU.

It is clear from figure \ref{fig:alfven hists} that the fraction of solar wind that is Alfv\'enic is much higher at 0.3 AU ($\sim 80\%$) as compared with 1 AU ($\sim 50\%$), which agrees well with previous studies \citep{Roberts1987a, Bruno2007}. At 0.3 AU the anisotropic population is almost always Alfv\'enic (ie.~$|\sigma_{c}| >$ 0.8), but this is not the case for the isotropic wind. We therefore propose splitting the solar wind at 0.3 AU into three populations based on the the observable boundaries in this parameter space:
\begin{itemize}
	\item An anisotropic, Alfv\'enic population
	\item An isotropic, Alfv\'enic population
	\item An isotropic, non-Alfv\'enic population
\end{itemize}
The split in anisotropy was chosen to be the saddle in between the two populations at $T_{p\perp} / T_{p\parallel} = 1.7$, and the split in cross-helicity was chosen at the edge of the Alfv\'enic population at $| \sigma_{c} | = 0.8$. These boundaries are shown in figure \ref{fig:alfven hists}. At 0.3 AU - 0.4 AU, $\sim$80\% of the wind was Alfv\'enic, split equally between isotropic and anisotropic, and the remaining 20\% was non-Alfv\'enic.

\begin{figure}
	\includegraphics[width=\columnwidth]{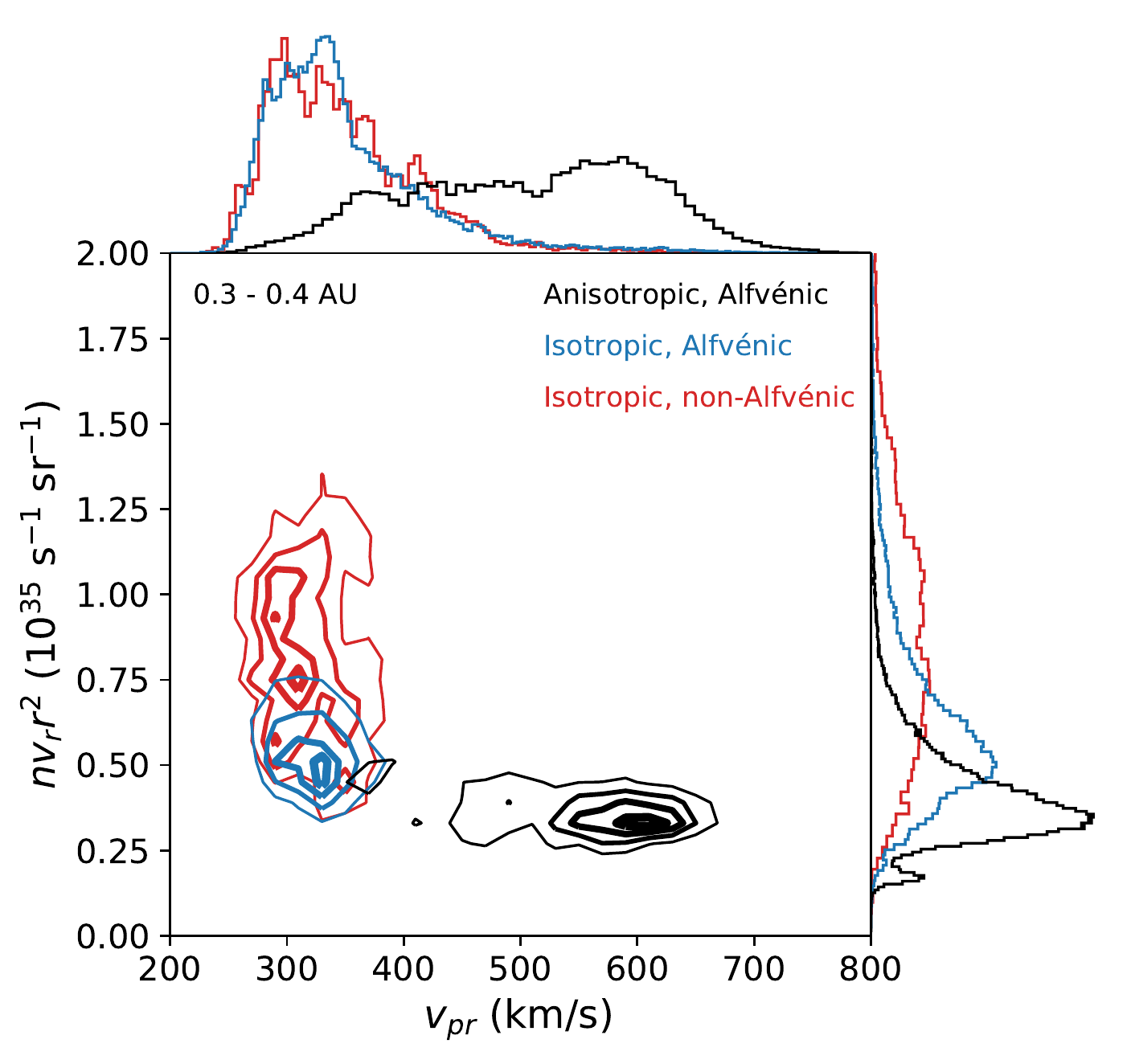}
	\caption{Contours of the 2D histograms of radial number density flux against radial velocity (centre panel) with adjoining 1D histograms of radial velocity (top panel) and radial number density flux (right panel) from data taken at 0.3 -- 0.4 AU. Different colours represent the three different categories of solar wind defined statistically in the top panel of figure 3. Contours are interpolated from a 2D histogram and plotted at levels of (0.3, 0.5, 0.7, 0.9) times the maximum bin value. 1D histograms are linearly scaled.}
	\label{fig:velocity hists}
\end{figure}

With this classification in mind, the top panel of figure \ref{fig:velocity hists} shows the radial velocity distribution of the solar wind in each category. Both isotropic populations consist primarily of solar wind with speeds less than 500 km/s, whereas the anisotropic population spans a wide range of speeds from 300 km/s -- 700 km/s. In fact, at 0.3 -- 0.4 AU Helios measured slightly more anisotropic solar wind below 500 km/s than above. This reinforces the idea that the concept of fast and slow solar winds breaks down at intermediate speeds where wind can have properties similar to either the very slow or very fast wind, and again suggests that some slow wind may have the same origin as fast wind.
Another known property of the fast solar wind is that the radial mass flux does not depend on speed \citep{Feldman1978, Wang2010}. To investigate whether this is true for the anisotropic wind as a whole, the main panel of figure \ref{fig:velocity hists} shows radial flux as a function of radial velocity for each of the three categories. The anisotropic wind has a constant flux that does not depend on speed; when evaluated at 1 AU this average flux is $\sim$~1.8~$\times~10^{8}$ cm$^{-2}$ s$^{-1}$, which agrees well with independent measurements made at 1 AU and beyond \citep{Phillips1995, Goldstein1996, Wang2010}. The isotropic Alfv\'enic wind has a slightly higher flux, whereas the non-Alfv\'enic wind has widely ranging fluxes varying up to 1 to 4 times the base value of the anisotropic wind, suggesting a distinct physical origin.

\begin{figure}
	\includegraphics[width=\columnwidth]{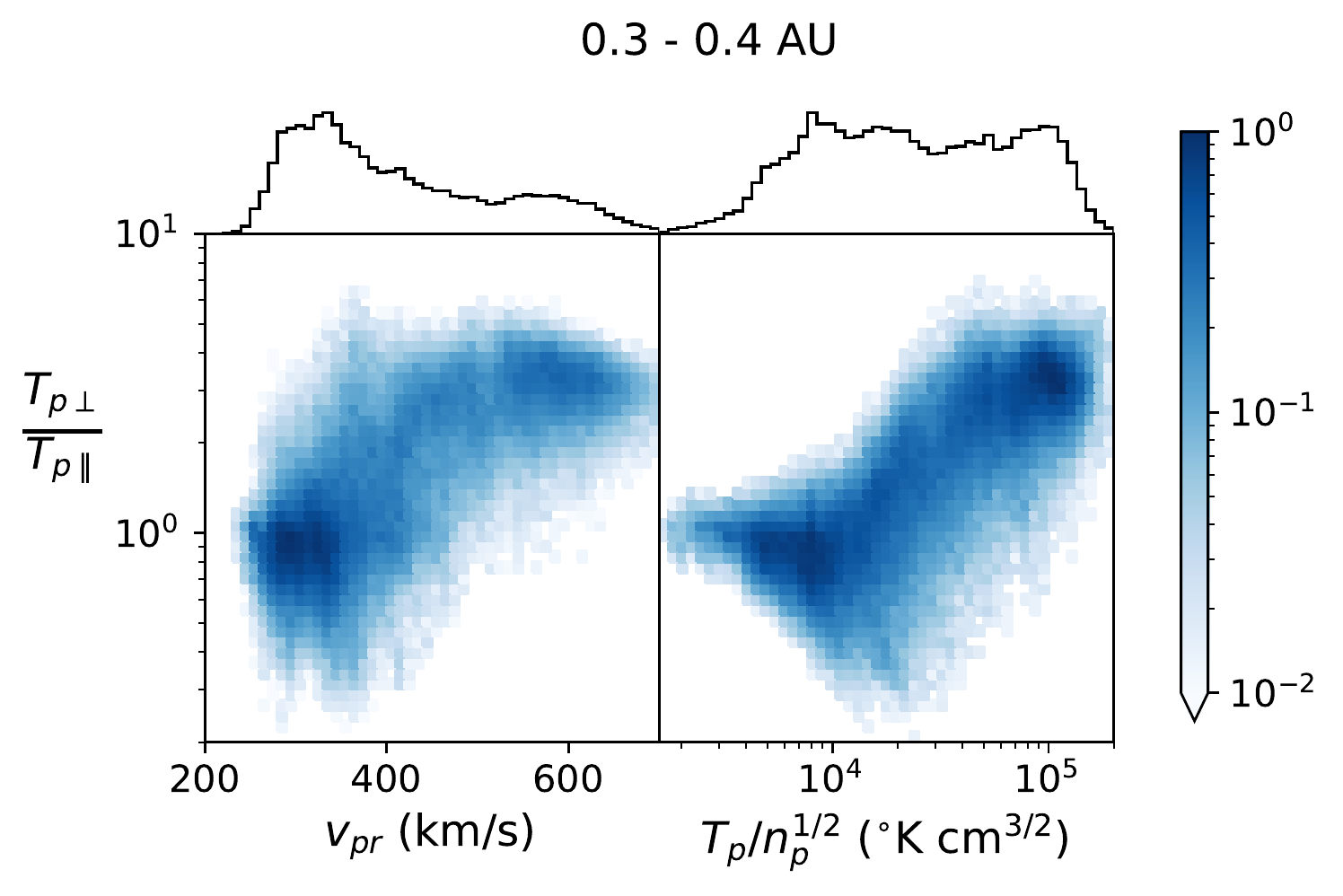}
	\caption{Joint probability distributions of temperature anisotropy (y--axis) against radial velocity (x--axis, LH panel) and proton specific entropy (x--axis, RH panel). 1D histograms for the x-axis variables are shown above the 2D histograms on a linear scale.}
	\label{fig:entropy hists}
\end{figure}

To investigate the link between the three categories and their source regions on the Sun we also looked at the dependence of temperature anisotropy on proton specific entropy (used later as a proxy for composition). Figure \ref{fig:entropy hists} shows the joint distribution of temperature anisotropy, and solar wind speed or proton specific entropy. At 0.3 AU, low speed wind (200 km/s -- 300 km/s) is all isotropic, and at high speeds (450 km/s -- 700 km/s) all of the wind is anisotropic, however, at intermediate speeds (300km/s -- 450 km/s) the wind is spread between the two different states of $T_{\perp} / T_{\parallel}$. In contrast the variation between temperature anisotropy and specific entropy is slightly smoother; isotropic wind corresponds exclusively to low entropy and anisotropic wind exclusively to high entropy, with a continuous variation in between. This result is used only as a correlation, and we do not claim that there is any causal relationship between entropy and temperature anisotropy. In section \ref{sec:composition} we discuss how this correlation can be used as an intermediate step to infer the compositional properties of our three different categories.

%%%
\subsection{Spatial distribution of the three solar wind populations}
\label{sec:timeseries}

\begin{figure*}
	\includegraphics[width=2\columnwidth]{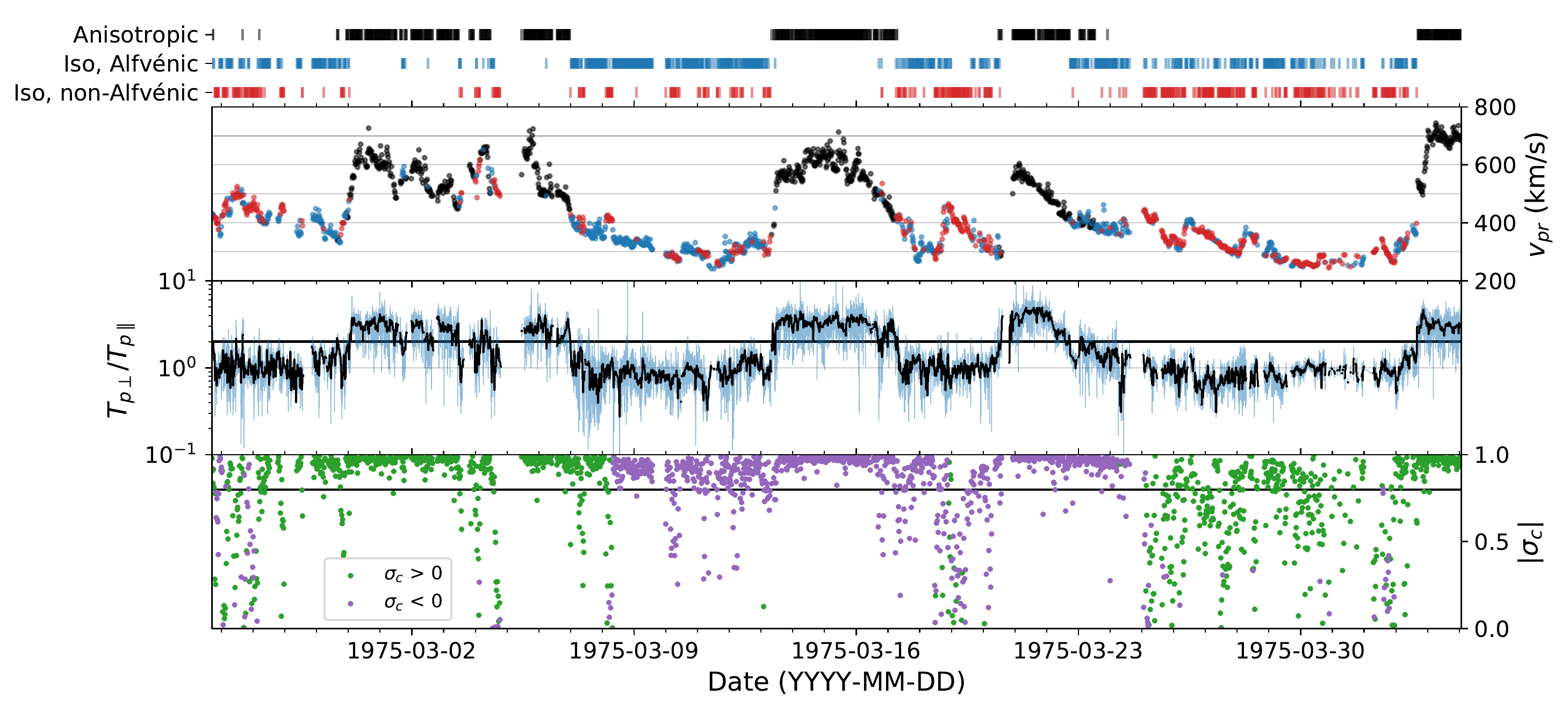}
	\caption{Timeseries data from the first perihelion pass of Helios 1, illustrating the categorisation displayed statistically in the top panel of figure \ref{fig:alfven hists}. Top panel shows the categorisation. Second panel shows the radial proton speed, coloured by the categorisation. Third panel shows full resolution proton temperature anisotropy (light blue) and 20 minute averaged values (dark black). The line at $T_{p\perp} / T_{p\parallel} = 1.7$ shows the boundary between anisotropic (points above) and isotropic (points below) wind. Bottom panel shows absolute values of 20-minute cross helicity, coloured by sign. The line at $\left | \sigma_{c} \right | = 0.8$ shows the threshold separating Alfv\'enic (points above) and non-Alfv\'enic (points below) wind.}
	\label{fig:timeseries}
\end{figure*}

We have shown that at 0.3 AU it is possible to distinguish between three types of solar wind based on the statistics of proton temperature anisotropy and Alfv\'enicity. To understand the spatial distribution of each population within the solar wind, figure \ref{fig:timeseries} shows the time-series measurements made by Helios 1 inside 0.5 AU during its first perihelion pass.

The bi-modal nature of proton temperature anisotropy is clear, even within the un-averaged and noisy 40.5 second cadence measurements, and the wind remained in one anisotropy state for days at a time. In contrast, within the isotropic category the Alfv\'enic and non-Alfv\'enic sub categories are well mixed and interspersed within each other. During some isotropic periods (e.g.~around 1975-03-09) the non-Alfv\'enic wind is sub-dominant and appears to be embedded in the Alfv\'enic wind, whereas at other times (e.g.~around 1975-03-30) there appears to be an approximately even mix of Alfv\'enic and non-Alfv\'enic wind.

The transition from isotropic to anisotropic wind was sharp, and always occurred at the leading edge of high speed streams. It is known that composition and entropy undergo sharp changes at the leading edge of high speed streams \citep{Wimmer-Schweingruber1997, Lazarus2003, Crooker2012a}, but figure \ref{fig:timeseries} demonstrates that a coincident temperature anisotropy boundary is also present. The sharp increases in temperature anisotropy were driven by increases in $T_{\perp}$ whilst $T_{\parallel}$ stayed roughly constant across the boundaries (not shown). The transition from anisotropic back to isotropic was also sharp, caused by sharp decreases in $T_{\perp}$, and occurred inside the rarefaction edge of high speed streams. The sudden drop in $T_{\perp}$, which caused a coincident drop in total temperature and therefore a drop in specific entropy, was not correlated with changes in the cross helicity. The only other observable change in the plasma and magnetic field data are magnetic field fluctuations that look qualitatively different on either side of the boundary (not shown). The time-series gives a clear visual demonstration that it is impossible to cut the velocity in a single place to separate different types of wind, but clear bi-modality makes a cut in temperature anisotropy easy. We re-iterate that performing this separation is only possible at heliocentric distances $<$ 0.8 AU, as at large distances the anisotropic wind becomes more isotropic, such that the two populations are no longer separable.

%%%%
\section{Linking in-situ measurements to solar sources}
\label{sec:solar sources}
We now use the observations made in section \ref{sec:results} to link our three categories of solar wind to their solar sources. A summary of the conclusions drawn in this section is given in table \ref{tab:summary}.

\subsection{Known properties of coronal hole wind}
It has long been known that wind originating on open field lines rooted inside large coronal holes forms the fast solar wind \citep{Krieger1973, Sheeley1976, Cranmer2009}. Remote sensing measurements also show pronounced temperature anisotropies present above coronal holes whilst the solar wind is still near to the Sun \citep{Kohl1997a, Cranmer2008}. Because the anisotropic category is the only one with high speeds (figure \ref{fig:velocity hists}), we infer that wind produced in the core of coronal holes belongs to our anisotropic category. The spatial distribution of anisotropic wind, with slower speeds always occurring in the rarefaction edges of high speed streams (figure \ref{fig:timeseries}) shows that rarefaction during transit is responsible for the relatively low speed of some anisotropic wind \citep{Pizzo1991}. The reason slower speeds are not observable at the leading edge of high speed streams is because by 0.3 AU they have already been accelerated by the faster wind to form a co-rotating interaction region \citep{Burlaga1974, Pizzo1991, McGregor2011, Richardson2018}.

Note that we have chosen to distinguish between the edges and the core of coronal holes; at the edge of coronal holes the magnetic field lines typically undergo large separations as a function of height in the corona, which has the effect of reducing both the wind speed \citep{Levine1977, Wang1991, Cranmer2007a, Pinto2016} and charge state ratios \citep{Wang2009}. In the next two sections further evidence is used to predict which one of our three categories coronal hole edge wind is part of.

\begin{table*}
	\caption{Properties of our three categories of solar wind near the Sun at solar minimum. The first 6 rows show properties directly measured by Helios at distances 0.3 AU -- 0.4 AU in this study. The bottom 4 rows show inferred properties. See section \ref{sec:solar sources} for more details.}
	\begin{tabular}{cccc}
		\hline
										&Isotropic non-Alfv\'enic	& Isotropic Alfv\'enic		& Anisotropic \\ \hline
		Fraction at 0.3 AU -- 0.4 AU			& 21.6 \%				&  37.4\%				& 39.0 \%		\\
		Speed							& 200 km/s -- 500 km/s	& 200 km/s -- 500 km/s	& 300 km/s -- 700 km/s  \\
		$T_{p\parallel}$						& 0.02 -- 0.4 MK 		& 0.02 -- 0.4 MK 		& 0.03 -- 0.4 MK	\\
		$T_{p\perp}$						& 0.01 -- 0.1 M K		& 0.01 -- 0.1 M K		& 0.1 -- 1 MK \\
		Entropy 							& Low				& Low				& High \\
		Mass flux							& Variable				& Constant			& Constant	\\ \hline
		Coronal freeze in temperature			& High				& High				& Low \\
		O$^{7+}$/O$^{6+}$					& High				& High				& Low \\
		C$^{6+}$/C$^{5+}$					& High				& High				& Low \\
		Solar source(s)						& Small scale transients	& Active regions, coronal hole edges		& Coronal hole cores	 \\
		\hline
	\end{tabular}
	\label{tab:summary}
\end{table*}

\subsection{Correlation of anisotropy, entropy, and composition}
\label{sec:composition}
At distances beyond 1 AU the proton specific entropy is anti-correlated with the O$^{7+}$/O$^{6+}$ charge state ratio \citep{Pagel2004}. In addition observations at 1 AU show that specific entropy is anti-correlated with the C$^{6+}$/C$^{4+}$ charge state ratio \citep{Stakhiv2016}. We have shown in figure \ref{fig:entropy hists} that entropy has a monotonic dependence on proton temperature anisotropy (but note again that this is not necessarily a causal relationship). Linking this newly observed relationship to the inferred relationship between entropy and charge state ratios suggests that anisotropic wind has low O$^{7+}$/O$^{6+}$ and C$^{6+}$/C$^{4+}$ charge state ratios, and isotropic wind has high charge state ratios. As well as being related statistically, the sharp boundaries between anisotropic wind and isotropic wind mimic the locations of sharp composition boundaries found in other studies (see section \ref{sec:timeseries} for a discussion). This backs up the statistical inference derived between proton temperature anisotropy and heavy ion charge states.

Using specific entropy as a bridge between anisotropy and composition therefore corroborates our previous conclusion that anisotropic wind originates in the core of large coronal holes, which are known to emit wind with low charge state ratios \citep{Geiss1995, Wang2009}. This allows us to infer that both categories of isotropic wind do not originate in the core of coronal holes, but may originate at coronal hole edges or outside coronal holes. This again agrees with remote sensing measurements that show reduced temperature anisotropies near the edges of coronal holes when compared to the core of coronal holes \citep{Susino2008}. We therefore suggest that the `isotropic' wind forms what is commonly thought of as the `slow solar wind'. There are a number of theories as to the origin of the slow solar wind \citep{Abbo2016}; in the next section we assign possible theories to either the Alfv\'enic or non-Alfv\'enic categories of the isotropic wind.

\subsection{Alfv\'enicity and mass flux variability}
Solar wind with a high Alfv\'enicity implies a steady state release of plasma on open field lines. This hypothesis is supported by the relatively constant mass flux in the Alfv\'enic isotropic wind (see figure \ref{fig:velocity hists}). This means any Alfv\'enic wind must have been released on field lines that remained open for at least the 20 minute resolution of the cross-helicity calculated from in-situ data. Areas of long lasting open field on the Sun can be split into the core of coronal holes (already categorised), edges of coronal holes, and active regions. Remote sensing measurements have shown that active region outflows have high coronal electron temperatures \citep{Neugebauer2002, Brooks2012}, contain open field lines allowing plasma to escape into the heliosphere \citep{Slemzin2013}, and can supply mass fluxes similar to those measured in-situ \citep{Brooks2015}. We therefore conclude active region wind is most consistent with the isotropic-Alfv\'enic category, along with the wind from the edges of coronal holes which also contain long lasting open magnetic fields and has similar properties.

Finally, we predict that the non-Alfv\'enic wind is consistent with the final known type of slow solar wind, typically called ``number density structures'' or ``blobs'', which have been detected both remotely \citep[e.g.][]{SheeleyJr.1997, Viall2015, DeForest2018} and in-situ \citep[e.g.][]{Kepko2003, Sanchez-Diaz2017a, Stansby2018d}. These are non steady state transient structures with a high density but similar speed as the surrounding slow wind, and therefore have enhanced mass fluxes relative to the background wind. This property is exactly what we have measured for the non-Alfv\'enic wind (figure \ref{fig:velocity hists}), backing up our final categorisation.

A summary of our mapping of possible solar sources to in-situ solar wind categories is given in table \ref{tab:summary}.

%%%%
\section{Comparison with other categorisation schemes}
\label{sec:compare}
Recently several authors have also attempted to categorise sources of the solar wind using in-situ observations, choosing the categories of coronal mass ejection wind, coronal hole wind, and interstream wind \citep[for a summary, see][]{Neugebauer2016}. In this paper we have deliberately removed coronal mass ejection wind from our dataset, and have used proton temperature anisotropy as the only variable separating coronal hole wind (anisotropic) and interstream wind (isotropic).

\cite{Zhao2009} used only the O$^{7+}$/O$^{6+}$ ratio and solar wind speed measured at 1 AU. This method is limited by the slow cadence (1 hour) of charge state ratio measurements available, but has the advantage that the O$^{7+}$/O$^{6+}$ ratio is known to be directly related to plasma properties near the Sun. Because the distribution of heavy ion charge states is only clearly bimodal at solar minimum \citep{Zurbuchen2002}, it is not clear if this method works well during solar maximum conditions. Because our method also assumes a bimodal distribution of charge state ratios, it is not clear if it is still applicable during solar maximum conditions either.

\cite{Xu2015} picked in-situ measurements, manually categorised specific intervals of the measurements, and then tried to find boundaries in a multi-dimensional parameter space that reliably split the data into the assumed categories. These boundaries could then be applied to other intervals where the categorisation is unknown. This method is practical and pragmatic for rapidly categorising solar wind sources, but the boundaries between categories are somewhat arbitrary and do not necessarily directly relate to the different physics of solar wind formation at each solar wind source. The advantage of the \cite{Xu2015} method is that it only uses single point measurements of solar wind protons and magnetic fields, so the cadence at which it can be applied is limited only by that of the in-situ measurements. In contrast our method is limited to a 20 minute cadence, which is in practice limited by the number of 40.5 second cadence of proton measurements needed to reliably calculate $\sigma_{c}$.

Other authors have backmapped solar wind measured at 1 AU to try and determine the exact location on the Sun from which it originated \citep[e.g.][]{Neugebauer1998, Fu2015, Fazakerley2016, Peleikis2017, Zhao2017}. This method assumes that the solar wind travels along magnetic field lines between the Sun's surface and magnetic field source surface at 2.5$r_{s}$ (solar radii), which are computed using a potential field source surface model, and then travels radially and at a constant speed to the in-situ observer. This method has the advantage of drawing a direct link by trying to predict the exact solar wind source location of in-situ measurements. Although it is successful in identifying sources on very large time scales of $\sim$ days, it is currently not possible to probe smaller scales, and does not take into account dynamical processing that occurs during transit between the Sun and the in-situ observer.

%%%%
\section{Conclusions and predictions for future missions}
\label{sec:predictions}
We have presented an attempt to map in-situ measurements of solar wind to their sources, using properties of the solar wind observable at 0.3 AU that are un-observable at 1 AU due to dynamical interactions during transit. We find that the solar wind can be split into three categories (summarised in table \ref{tab:summary}), based on in-situ measurements of proton temperature anisotropy and Alfv\'enicity (section \ref{sec:results}), and sort possible solar origins of the solar wind into these three categories (section \ref{sec:solar sources}). Although many other methods have been developed to attempt the same goal of solar source categorisation (section \ref{sec:compare}), the lack of in-situ composition and remote sensing data available during the Helios era (1974 -- 1984) restricted our ability to use these more modern techniques. However, in the near future we will have access to simultaneous in-situ measurements of protons in the inner heliosphere, in-situ measurements of solar wind composition, and a wide range of remote sensing data. We finish by describing how our new categorisation scheme can be applied to data from upcoming missions to the inner heliosphere.

Parker Solar Probe \citep[PSP,][]{Fox2015} will make in-situ measurements of the solar wind at heliocentric distances inside 0.3 AU, and the first comprehensive in-situ solar wind measurements inside 1 AU since Helios. Proton and magnetic field measurements made by PSP will allow us to perform the categorisation scheme outlined in this paper. Advances in modelling and remote sensing since the Helios era mean that it will then be possible to backmap the solar wind measured by PSP to a predicted source location on the Sun. If our categorisation is correct, the three categories of in-situ solar wind will backmap to their respective inferred solar sources.

Solar Orbiter \citep[SO,][]{Muller2013} will provide the first solar wind composition measurements between 0.3 AU and 1 AU. This will allow us to directly test the correlation between temperature anisotropy and charge state ratios, without having to bridge the gap by using proton specific entropy as an intermediate variable. If our categorisation is correct, isotropic wind will clearly correspond to high charge state ratios, and anisotropic wind will clearly correspond to low charge state ratios. In addition SO will carry on board remote sensing instruments that are designed to target the predicted solar sources of in-situ measurements, making backmapping wind to its source even more accurate than using remote sensing instruments at 1 AU.

\section*{Acknowledgements}
D.~Stansby is supported by STFC studentship ST/N504336/1, and thanks Trevor Bowen, Allan MacNeil, Denise Perrone and Alexis Roulliard for helpful discussions. T.~S.~Horbury  is  supported  by STFC grant ST/N000692/1. This work was supported by the Programme National PNST of CNRS/INSU co-funded by CNES.

Data were retrieved using HelioPy v0.5.3 \citep{Stansby2018f} and processed using astropy v3.0.3 \citep{Price-Whelan2018}. Figures were produced using Matplotlib v2.2.2 \citep{Hunter2007, Droettboom2018}.

Code to reproduce the figures presented in this paper is available at \url{https://github.com/dstansby/publication-code}.

%\newpage
%~
%\newpage
\bibliographystyle{mnras}
\bibliography{/Users/dstansby/Dropbox/Physics/library}

\end{document}